\documentstyle[prl,aps,epsfig]{revtex}
\newcommand{\str}{\mathop{\rm str}\nolimits}

\newcommand{\cD}{{\cal D}}
\newcommand{\cO}{{\cal O}}
\newcommand{\cN}{{\cal N}}

\renewcommand{\vr}{{\bf r}}

\newcommand{\vp}{{\bf p}}
\newcommand{\vn}{{\bf n}}

\newcommand{\vnabla}{{\bf \nabla}}

\begin{document}

\author{ V. Tripathi$^{1}$ and D.E.Khmelnitskii$^{1,2}$}
\address{Cavendish Laboratory, University of Cambridge, Madingley Road,
Cambridge, CB3 0HE, UK$^{1}$ \\and L.D.Landau Institute for Theoretical 
Physics, Moscow, Russia$^{2}$} 
\title{Level statistics in a two-dimensional disc with diffusive boundary
scattering}  

\date{\today}

\maketitle

\begin{abstract}

We calculate the energy level statistics in a two-dimensional disc with
diffusive boundary scattering by the means of the recently proposed ballistic
nonlinear $\sigma$-model \cite{MK95}.   

\end{abstract}
\bigskip
\bigskip


$\bf 1$. The purpose of this paper is to use the recently proposed
nonlinear $\sigma$-model \cite{MK95} for ballistics in disordered
conductors with long mean free path and find out how it works. 
In a recent paper \cite{AA} Andreev and Altshuler (AA)
suggested a general method of calculation of level statistics in a
disordered system beyond the limits of the random matrix theory. Their
calculations were performed for a diffusive 
disordered system with the mean free path $l$ shorter than the system
size $R$. For this case and for energy difference $\omega$ exceeding
the mean level spacing $\Delta$ ($\omega \gg \Delta$) their method was 
based on using the non-linear $\sigma$-model \cite{Efetov} and
accounting for a perturbative contribution from the vicinity of several
stationary points of the action \cite{Efetov}. AA have also 
conjectured a general form which the level statistics obeys. 
In this paper we follow the same general strategy, addressing this 
problem for a quantum particle in a two-dimensional disc with no 
scattering in the bulk and strong boundary scattering\cite{dif}. Such 
a problem is a natural target for the recently proposed field theory for
quantum ballistics \cite{MK95,AAAS}.     
Consideration is restricted solely to unitary symmetry. In all relevant
parts our results coincide with those independently obtained by Blanter,
Mirlin and Muzykantskii \cite{BMM2}. 

$\bf 2$. In order to apply the general approach of AA, we begin with
replacing the non-linear diffusive super-matrix 
$\sigma$-model \cite{Efetov} by its ballistic generalization
\cite{MK95}. The partition function of this field theory is determined as
a functional integral over a supermatrix $g(\vn, \vr) = U^{-1} \Lambda U$
on the energy shell $E = \vp^2/2m, \; (\vn = \vp/p)$ in the phase space: 
\begin{equation}
Z = \int_{g^2=1} \cD g \exp\{-F\};  
\nonumber
\end{equation}
\begin{equation}
F =  \frac{\pi \nu}{4} \str \int d \vr \left[i\omega \Lambda \langle g(\vr)
\rangle - 2v_F \langle \Lambda U^{-1} \vn \vnabla U \rangle  
+ \int d \cO d \cO' W_{\vn,\vn'} g(\vr,\vn) g(\vr,\vn') \right],   
\label{Z-F}
\end{equation}
where $\langle ..\rangle =  \int ...  d \cO /2\pi, \; d \cO = d \cO_{\vn},
d \cO' = d \cO_{\vn'}$ and the scattering probability $W(\vn, \vn')$ in
the bulk is connected with the mean free time $\tau$ and transport mean
free time $\tau_{\mbox{tr}}$ as 
\begin{equation}
\label{tau}
\frac{1}{\tau} = \int d \cO W(\vn, \vn');  \;
\frac{1}{\tau_{\mbox{tr}}}  = \int d \cO W(\vn, \vn') \left( 1 - \vn \vn'
\right). 
\end{equation}  
In this paper we will consider a clean disc with no scattering in the
bulk and strong boundary scattering. Therefore, both   $\tau$ and
$\tau_{\mbox{tr}}$ will be taken as being infinitely large.  
Integration in Eq~(\ref{Z-F}) is not defined unless boundary conditions
are imposed on super-matrix $g$ at the inner boundary of the sample. 
Since super-matrix $g$ has a meaning of a distribution function of
electrons, the boundary condition it obeys is similar to that, which is
applied to the distribution function in classical kinetics \cite{LP}
\cite{Fuchs}. General boundary condition for matrix-functions (see
\cite{Ovchin} for an example) is pretty complicated. Fortunately, for the
purposes of this paper (calculation of the spectral correlation function
with the precision to the first non-vanishing term beyond Random matrix
theory) the problem could be significantly simplified, because most of
the properties of the energy levels could be determined by the the values
of $g$-matrix close to the special point $g(\vr \vn) = \Lambda$. If $U =
1 - w/2 + w^2/8 + .. $, then the free energy $F$ could be expressed
through matrices $w$, what gives in the quadratic approximation:
\begin{equation}
F_0 \{\hat w\} = -\frac{\pi \nu}{4} \int d\vr d\cO_{\vn} \str [w_{21}
(\hat L - i\omega) w_{12} ], 
\label{quadratic}
\end{equation}
where indices $1$ ($2$) relate to "retarded" ("advanced") degrees of
freedom, and $\hat L$ denotes operator of the kinetic equation. Since the
free energy in Eq~(\ref{quadratic}) is quadratic, it results in classical
linear equation.  The
boundary condition which should be imposed upon $w_{12} (\vr, \vn)$, is
now a direct analog of the condition, imposed upon the distribution
function in classical kinetics. Extremely strong boundary scattering is
popularly modelled by the diffusive boundary condition \cite{LP},
\cite{Fuchs}, which assumes that the distribution function for outgoing
particle does not depend on angular variable $\vn$ and is coupled to that
for incoming particle by flux conservation. If $\cN$ is an outward
normal to the sample's boundary, then the diffusive boundary condition
reads as 
\begin{equation}
\label{Fuchs}
w_{12} (\vn \cN < 0) = \int_{\vn' \cN > 0} \frac{d\cO_{\vn'}}{\pi} \vn \cN
w_{12}(\vn').  
\end{equation}  

$\bf 3$.  According to AA, the level statistics is determined by the
determinant of a linear operator $\hat L$ from Eq~(\ref{quadratic}). The
eigenvalue condition is 
\begin{equation} 
\label{eigen1} 
\vn v_F \vnabla w = \lambda w,
\end{equation}
subject to boundary condition
\begin{equation} 
\label{Fuchs-1} 
2 w_< =- w_< \int_{\pi/2}^{3\pi/2} d\phi \cos \phi = \int_{-\pi/2}^{\pi/2}
d\phi \cos \phi  w_> (\phi),
\end{equation}
where $w<$ and $w_>$ are the values of "distribution function" $w(\vn)$ at
the disc boundary at $\vn \cN > 0 $ and $\vn \cN < 0$ respectively and
$w<$ does not depend on its argument. 

The left hand side of Eq~(\ref{eigen1}) consists of a derivative $\partial
w/\partial l$  along the trajectory of a particle inside the disc (see
Fig 1).  Its solution has the form of a simple exponential  
\begin{equation} 
\label{eigen2} 
w(l) = w(0) \exp \left[\frac{\lambda l}{v_F} \right]
\end{equation}
Solution (\ref{eigen2}) should be substituted into the boundary condition
(\ref{Fuchs-1}). It is also convenient to express the direction of
momentum $\cos \phi$ of incident electron at point $\theta$ in of the
disc boundary Eq~(\ref{Fuchs-1}) through that coordinate on the boundary
$\theta'$, where this electron was diffusively scattered from ($\cos \phi =
\sin [( \theta' - \theta)/2]$) . This all leads to the eigenvalue
equation in the form:
\begin{equation} 
\label{eigen3} 
4w_<(\theta) = \int_{\theta}^{\theta + 2\pi}   \exp \left[\frac{2\lambda
R}{v_F}\sin \left(\frac{\theta' - \theta}{2}\right) \right] \sin
\left(\frac{\theta' - \theta}{2}\right) w_< (\theta') d\theta'.  
\end{equation}
The expansion of $w_<(\theta)$ in the Fourrier series $w_<(\theta) = \sum w_m
e^{im\theta}$ transforms the condition (\ref{eigen3}) into 
\begin{equation} f_m (\mu_{m,k}) = 0, \quad  
f_m (\mu) = 1 - \frac{1}{2} \int_0^{\pi}  \exp \left[2imu +  \mu \sin u
\right] \sin u du =   0,
\label{eigen4} 
\end{equation}
where $\mu_{m,k} = 2 R \lambda_{m,k}/v_F$. One can see from
Eq~(\ref{eigen4}) that one of the eigenvalues with $m=0$ vanishes  ( say
$\mu_{0,0} = 0$). This corresponds to $w$ independent of both $\vn$ and
$\vr$ and it is not surprising that the relaxation rate of this
eigen-mode vanishes. Substitution $m \rightarrow
-m$  into Eq~(\ref{eigen4}) makes it clear that $\mu_m =
\mu_{-m}$. The equation, complex conjugate to Eq~(\ref{eigen4}) shows
that if $\mu$ is an eigenvalue, then $\mu^*$  is an eigenvalue as
well. None of the eigenvalues has a negative real part. A natural
labeling \cite{BMM2} is $k = 0, \pm 1, \pm 2, ...$ for even $m$ and $k =
\pm 1/2, \pm 3/2 ..$ for odd $m$. For $k=0$ and even  $m$ the eigenvalues
are real.  The asymptote of the eigenvalues is
\begin{equation}
\label{asymt}
\mu_{m,k} \approx \frac{\ln k}{4} + \pi i \left(k + \frac{1}{8}\right ) 
\quad 0 \le m \ll k.
\end{equation}
So, for $ 0 \le m \ll k$  $\mbox{Im} \mu \gg \mbox{Re} \mu$ and both don't
depend on $m$. (Some the eigenvalues of the Liouvillean are shown in Fig 2.) 

{\bf 4}. The purpose of this paper is to calculate  the
spectral correlation function 
\begin{equation}
\label{R2}
R_2(\omega) = (\pi \Delta R^2)^2 \langle \nu(\epsilon) \nu(\epsilon +
\omega )\rangle  - 1,  
\end{equation}
where $\nu(\epsilon)$ is the density of states and $\Delta = 1/\pi R^2
\nu$ is the  mean level spacing. The time of ballistic flight along
diameter of the disc $t_f = 2 R/v_F$ introduces a natural scale for the
frequencies. \\
As has been shown by AA,  the deviation of $R_2(\omega)$ from the
Wigner-Dyson expression   
\begin{equation}
\label{RMT}
R_2(s) = \delta (s) - \frac{\sin^2 \pi s}{s^2}, 
\quad s = \frac{\omega}{\Delta} 
\end{equation} 
at frequencies  $\omega \gg \Delta$ is well described by introducing the
spectral determinant $D(s)$ 
\begin{equation}
\label{det}
D(s) =  \prod_{m,k \ne (0,0)}
\frac{\lambda_{k,m}^2}{(\lambda_{m,k} - is\Delta)(\lambda_{m,k} +
is\Delta)}, 
\end{equation} 
which is closely connected with the spectral function $S(\omega)$, first
introduced by Altshuler and Shklovskii \cite{AS} for diffusive systems:
\begin{equation}
\label{AS}
S(\omega) =  \sum_{m} \sum_{k} (\lambda_{m,k} - i\omega)^{-2} ;   
\quad \frac{\partial^2 \ln D(s)}{\partial s^2}  = -2 \left( \Delta^2
\mbox{Re} S(s\Delta) + \frac{1}{s^2}\right). 
\end{equation}
The spectral correlation function can be decoupled at $\omega \gg \Delta$
into the sum \cite{AA} of a smooth part $R_{sm}$\cite{AS} and an oscillating part
$R_{osc}$:   
\begin{eqnarray}
\label{AA-1}
R_{sm} (s) = \frac{\Delta^2}{2\pi^2} \mbox{Re} S(s\Delta) \\
\label{AA-2}
R_{osc} (s) = \frac{1}{2\pi^2 s^2 } D(s) \cos 2\pi s .
\end{eqnarray}
So, the calculation of the spectral determinant $D(s)$ is a key point of the whole
AA programme, which we approach now.

{\bf 5}.  It is possible to write down an expression for the spectral
determinant without an explicit computation of the eigenvalues. In order
to   
do that, note that the function $f_m(\mu)$, defined by Eq~(\ref{eigen4}),
is an entire function of its argument, which has only simple zeros at
$\mu = \mu_{m,k} $ and $f'(\mu)/(\mu\,f(\mu)) $ vanishes as 
$\mu \rightarrow \infty $.  Therefore, $f_m(\mu)$  can be represented as 
an infinite product ($m \ne 0 $). 
\begin{eqnarray}
\nonumber
f_m (\mu)=f_m (0)\,\exp \left[ \frac{f'_m (0) \mu}{f_m(0)} \right] \,
\prod_{k} \left[1 -  \frac{\mu}{\mu_{m,k}} \right] \, \exp \left\{\frac{\mu}{\mu_{m,k}}\right\} =\\
 = \frac{4m^{2}}{4m^{2}-1} \, \exp \left[\frac{f'_{m}(0)\mu}
{f_{m}(0)}\right] \,\prod_{k}\left[1-\frac{\mu}{\mu_{m,k}}\right]  \exp
\left\{\frac{\mu}{\mu_{m,k}}\right\}. 
\label{Wei}
\end{eqnarray}
For $m=0 $  the function $f_0 (\mu) $ vanishes at $\mu=0$.  So the same
theorem could be applied to the function $- 4 f_{0}(\mu)/ \pi\mu $.
Multiplying $f_{m}(\mu) $ and $f_{m}(-\mu) $, taking the product over all
$m$, and analytically continuing to $\mu = \pm i\xi =\pm i \omega t_f$,
we arrive, finally, at the expression for the spectral determinant
\begin{equation} 
D (\xi)  =  \frac{ \xi^2}{4} \left(\frac{\pi}{2}\right)^6 \prod_{m=-\infty}^{+\infty}[f_{m}(i\xi)\,f_{m}(-i\xi)]^{-1},
\label{det1} 
\end{equation}
where it is taken into account that 
\begin{equation}
\prod_{m=1}^{\infty} \frac{4m^{2}-1}{4m^{2}} = \left(
\frac{2}{\pi z} \sin \frac{\pi z}{2} \right)_{z=1} = \frac{2}{\pi}.
\label{det2}
\end{equation}
Since $\lambda_{m,k} = \mu_{m,k}  / t_f$, the spectral determinant $D(s)$
consists of two dimensionless parameters $\omega t_f$ and $\Delta t_f$.
One of this parameters is always small ($\Delta t_f \ll 1 $), while the
second one $\omega t_f$ could be either larger or smaller than
unity. These two limiting cases constitute the limits of high and small
frequencies respectively.  

{\bf 6}. At low frequencies $\omega t_f \ll 1$ the spectral determinant
$D(\xi)$ can be simlified and the asymptotes of both the smooth and the
oscillatory parts of the spectral correlation functions coincide, as was
first discovered by Kravtsov and Mirlin\cite{KM}. This gives the 
following expression for the spectral correlation function
\begin{equation}
\label{KM}
R_2 (s) = \delta(s)  -\frac{\sin^2 \pi s}{\pi^2 s^2} + B \frac{\Delta^2
t_f^2}{\pi^2} \sin^2 \pi s, \quad  B =  \sum_{m,k \ne 0,0}
\frac{1}{\mu_{k,m}^2}.   
\end{equation} 
Using the low frequency asymptote of Eq~(\ref{det1}), we can present the
spectral function 
$$
S(\omega) = \sum_m S_m(\omega)
$$ 
in the form of a contour integral 
\begin{equation}
\label{contint}
S_m (\omega) = \frac{t_f^2}{2\pi i} \oint_C \frac{dz}{(-i\omega t_f +z)^2}
\frac{d \ln f_m (z)} {dz} = - t_f^2 \, \left[\frac{d^2}{dz^2} \ln f_m(z)
\right]_{i\omega t_f},   
\end{equation}
where contour $C$ encloses all zeros of of the function $f_m(z)$. 
As $\omega \to 0$ we obtain from Eq~(\ref{contint}) the following
expression for the coefficient $B$ in Eq~(\ref{KM})
\begin{equation}
\label{B}
B = - \frac{19}{27} - \frac{175 \pi^2}{1152}  + \frac{64}{9\pi^2} \approx
- 1.48
\end{equation}

{\bf 7}. In order to find the asymptote of the spectral determinant $D$
in the high frequency limit, consider a product $P(i \xi)$ 
 \begin{equation}
 P(i\xi) = \prod_{m = -\infty}^{+\infty} f_m (i\xi).
 \label{P}
 \end{equation}
Its logarithm is presented by the sum  
\begin{equation}
\ln P(i\xi) = \sum_{m= -\infty}^{+\infty} \ln\left[1-\frac{1}{2} \int_0^{\pi} du
\sin u \exp (2im u + i\xi \sin u) \right]
\label{log}
\end{equation}
At this tage, it is convenient to use the indentity
\begin{equation}
\sum_{m= -\infty}^{+\infty} F(e^{2imu}) =\int_{-\infty}^{+{\infty}
}dx  \sum_{n= -\infty}^{+\infty} e^{2i \pi n x}  F(e^{2ixu}),
\label{ind}
\end{equation}
which replaces the sum over $m$ by the sum over $n$ and integral over
$x$. For large values of $\xi $ the integral over $u$ in Eq(\ref{log}) is
small and the logarithm should be expanded up to the second order in this
integral (linear term vanishes). After calculation the sum over $n$ the
expression could be simlified to the following form:   
\begin{equation}
\ln  P(i\xi) \approx   -  \frac{\pi}{8}  \int_0^{\pi}
du \sin^2 u \, \,\exp ( 2i\xi \sin u)  .
\label{simple}
\end{equation}
To evaluate the spectral determinant, we need to find the product $
P(i\xi)  P(-i\xi)$ Using the steepest descent method, we arrive at  $\xi
\gg 1$ at following asymptote  for the spectral determinant:
\begin{equation}
D(\xi) \approx \frac{\xi^2}{8}  \left(1 + \frac{\pi}{4}
\sqrt{\frac{\pi}{ \xi}} \cos\left(2 \xi - \frac{\pi}{4} \right) \right).
\label{spec-1}
\end{equation} 
This gives the smooth part of the spectral correlation function in the
form:  
\begin{equation}
\label{smooth-1}
R_{sm} (\omega) = \frac{ \Delta^2 t_f^{3/2}}{4 \sqrt{ \pi
\omega}} \cos \left(2 \omega t_f - \frac{\pi}{4} \right). 
\end{equation}
Eq~(\ref{AA-2}) gives the oscillatory part of the spectral function equal
to    
\begin{equation}
\label{osc-1}
R_{osc} (\omega) = \frac{\pi^4}{516} \,\, (\Delta t_f)^2 \, \cos
\left(\frac { 2\pi \omega}{\Delta} \right).  
\end{equation}
{\bf 8}. In conclusion, we found that the application of the ballistic 
non-linear $\sigma$-model\cite{MK95} to the study of level statistics 
for electrons in a clean disc with strong boundary scattering enables
us to solve this problem beyond the limits of the random matrix theory. 

A clean disc with diffusive scattering on its boundaries, unlike other
chaotic systems, has an upper limit for the time of flight at  $t = t_f
\equiv 2 R/v_F$.  Therefore, if a Fourier transform of a time dependnet
form-factor is calculated, it oscillates as a frequency  $\omega$  with
period $2\pi / t_f$. As it was shown in section 7, the smooth part of the
spectral corerelation function  $R_2 (\omega)$ at high frequencies
$\omega t_f \gg 1$ is proportional to square of the relevant form-factor.
This leads to oscillations of  $R_{sm} (\omega) $ with twice as shorter
period  $\pi / t_f$. In our understanding, such oscillations would not
appear in a general case.  

Another striking result is exhibited in Eq~(\ref{osc-1}): the amplitude
of the oscillatory part of spectral correlation function is small as
$(\Delta t_f)^2$, but does not decay with $\omega$, unlike one obtained
by AA for a diffusive system. This could be understood if to recall that
our disc is clean inside and, therefore, at short times $t \ll t_f$
certain correlation between electron wave function remains. This
correlation is small and proportional to $(p_F R)^{-2} \sim (\Delta
t_f)^2$, but decays with the energy difference $\omega$ much slower. If
our disc has a bulk disorder with the meen free time $\tau \gg t_f$
\cite{AlGef}, $R_{osc} \propto \exp[ - \omega \tau]$. To similar result
leads variation of the Fermi velocity with energy: $R_{osc} \propto \exp[
- \omega /E_F ]$.  The smooth part of spectral correlation function (see
Eq~(\ref{smooth-1}) ) also exhibits weak dependence on energy difference
$\omega$. In our understanding, these results are of a general nature.   

We are greatful to B.~A.~Muzykantskii, who participated in this 
work at its initial stage, for numerous conversations during the 
process of work and to Ya.~M.~Blanter and A.~D.~Mirlin and
B.~A.~Muzykantskii for the opportunity to read their paper \cite{BMM2}
prior to publication.


\newpage

\begin{figure}
\epsfxsize=10cm
\centerline{\epsffile{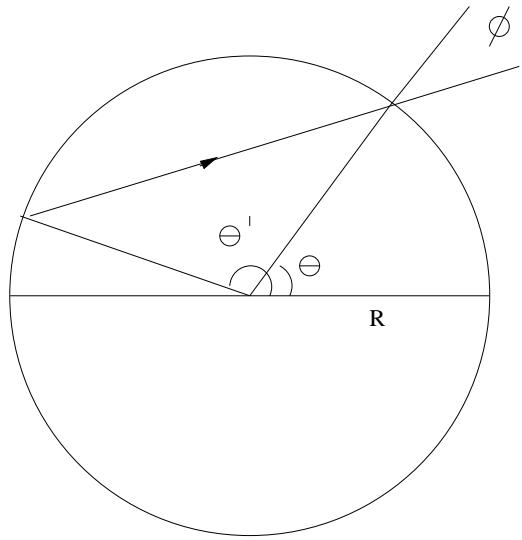}}
\caption{Typical electron trajectory}
\end{figure}

\bigskip
 
\begin{figure}
\epsfxsize=10cm
\centerline{\epsffile{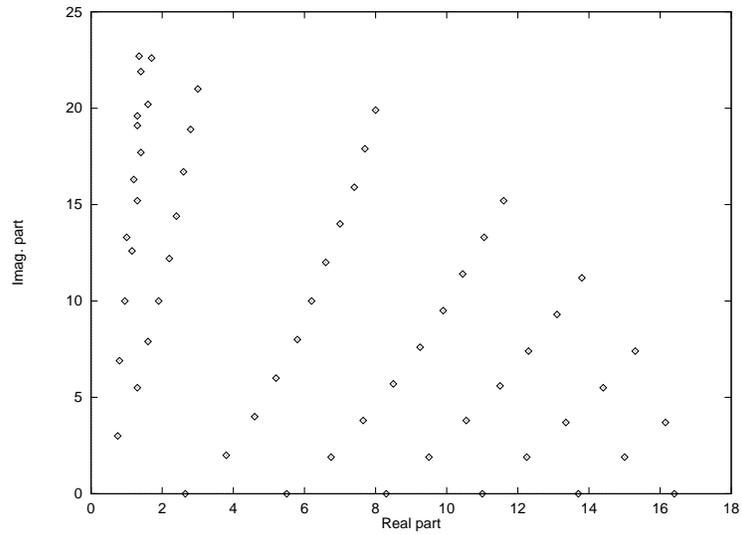}}
\caption{Eigenvalues of Liouvillean  operator in the disc with diffusive
boundaries.}
\end{figure}

\end{document}